\documentclass[lettersize,journal]{IEEEtran}
\usepackage{amsmath,amsfonts}
\usepackage{algorithm}
\usepackage{array}
\usepackage[caption=false,font=normalsize,labelfont=sf,textfont=sf]{subfig}
\usepackage{textcomp}
\usepackage{stfloats}
\usepackage{url}
\usepackage{verbatim}
\usepackage{graphicx}
\usepackage{cite}
\usepackage[noend]{algpseudocode}
\algdef{SE}[EVENT]{Event}{EndEvent}[1]{\textbf{upon event}\ #1\ \algorithmicdo}{\algorithmicend\ \textbf{event}}%
\algtext*{EndEvent}
\makeatletter
\newcommand{\algmargin}{\the\ALG@thistlm}   
\makeatother
\algnewcommand{\indentedState}[1]{\State%
    \parbox[t]{\dimexpr\linewidth-\algmargin}{\strut #1\strut}}

\usepackage{booktabs}
\usepackage{multirow}

\begin{document}

\title{A Deep Reinforcement Learning Approach for Cost Optimized Workflow Scheduling in Cloud Computing Environments}

\author{Amanda Jayanetti, Saman Halgamuge \IEEEmembership{Fellow, IEEE}, Rajkumar Buyya
\IEEEmembership{Fellow, IEEE}
\thanks{The authors are with Cloud Computing and Distributed Systems (CLOUDS) Lab, School of Computing and Information Systems, University of Melbourne, Melbourne, VIC 3010, Australia}}

\maketitle

\begin{abstract}
Cost optimization is a common goal of workflow schedulers operating in cloud computing environments. The use of spot instances is a potential means of achieving this goal, as they are offered by cloud providers at discounted prices compared to their on-demand counterparts in exchange for reduced reliability. This is due to the fact that spot instances are subjected to interruptions when spare computing capacity used for provisioning them is needed back owing to demand variations. Also, the prices of spot instances are not fixed as pricing is dependent on long term supply and demand. The possibility of interruptions and pricing variations associated with spot instances adds a layer of uncertainty to the general problem of workflow scheduling across cloud computing environments. These challenges need to be efficiently addressed for enjoying the cost savings achievable with the use of spot instances without compromising the underlying business requirements. To this end, in this paper we use Deep Reinforcement Learning for developing an autonomous agent capable of scheduling workflows in a cost efficient manner by using an intelligent mix of spot and on-demand instances. The proposed solution is implemented in the open source container native Argo workflow engine that is widely used for executing industrial workflows. The results of the experiments demonstrate that the proposed scheduling method is capable of outperforming the current benchmarks. 
\end{abstract}


\section{Introduction}

Cloud computing leverages virtualization techniques for providing users with convenient access to a pool of scalable resources \cite{BuyyaYVBB09}. As opposed to maintaining their own computing infrastructures, the pay-as-you-go model of cloud computing paradigm enables users to acquire a diverse range of virtual machines with varying flavors (CPU, Memory etc.) for meeting business needs in a more cost effective manner. The flavor of virtualized instances used for executing  tasks determines the total execution times of the workflows as well as the associated monetary costs. In order to maximize the achievable cost savings achievable while also ensuring the performance is maintained to a satisfactory level, it is imperative that cost optimized scheduling strategies are designed and implemented. 

In particular, the intelligent use of a mix of on-demand and spot instances for workflow executions is a potential means of achieving high cost efficiencies without adversely affecting performance expectations. Spot instances are offered by cloud providers at steep discounts compared to their on-demand counterparts in exchange for reduced reliability. This is because the cloud providers utilize spare computing capacities available for provisioning spot instances, and therefore when the capacity is needed back, the instances are interrupted. Furthermore,as opposed to on-demand instances with fixed prices, the prices of spot instances are not guaranteed to be fixed, as the pricing is dependent on long term supply and demand. The possibility of interruptions and pricing variations adds a layer of complexity that needs to be efficiently handled for enjoying the cost savings without compromising the underlying business requirements. Therefore, it is imperative to establish the right balance between the use of on-demand and spot instances for workflow executions in cloud computing environments. 

The ability of Reinforcement Learning (RL) agents to operate in stochastic environments, and learn through experience to act in an optimal manner amid highly dynamic conditions and uncertainties makes it an ideal candidate for overcoming the aforementioned challenges. While many heuristics and meta-heuristics have been proposed for cost optimized workflow scheduling, only very few works have explored the potential of RL in this area. In particular,  Deep Reinforcement Learning (DRL) \cite{DRLSurvey2024} has emerged as an efficient means of solving highly complex problems as evidenced by the recent successes achieved by DRL agents in complex control tasks in fields such as robotics, autonomous driving, healthcare and so on. In this work, we leverage the advanced capabilities of DRL for designing a cost optimized workflow scheduling framework. 

The design of action space is a fundamental characteristic of a DRL based formulation of a problem. The action spaces of a vast majority of scheduling problems that are modeled as DRL problems, include a flat set of actions. The action space may be discrete or continuous, and the agent selects an action from the action space. In this work, we propose a novel hierarchical way of designing the action space of the DRL model such that there is a clear distinction between on-demand and spot instances in action selection. A DRL framework comprising multiple actor networks guided by a common critic network is then designed to select a combination of actions from the hierarchical action space, to optimize cost of workflow executions. 

Container orchestration engines such as Kubernetes can seamlessly operate atop highly distributed and heterogeneous infrastructures and abstract away the complex coordination details from users. This in turn has enabled users to conveniently deploy workloads across a variety of cloud deployments ranging from private and public clouds to hybrid combinations of these. Complementary frameworks such as Argo workflow engine have emerged to extend the functionalities of Kubernetes to facilitate the management of more complex workloads such as Workflows. The schedulers of these frameworks are pre-configured to follow basic scheduling policies such as bin-packing. These simple policies are not capable of satisfying the complex cost optimization requirements of users. In order to achieve complex user-defined goals it is imperative to incorporate more advanced scheduling policies in the aforementioned workflow management engines. These policies should be capable of adapting to highly stochastic conditions that are inherent in clusters deployed in cloud computing environments. In this regard, we present an end-to-end means of training and deploying the DRL agent proposed in this work in the Argo workflow engine. 

More specifically, the following summarizes the main contributions of this work:

\begin{itemize}
    \item A DRL model for cost optimized scheduling of workflows in a cloud computing environment with the use of a balanced mix of on-demand and spot instances.
    \item A logical organization of the cluster in a hierarchical manner, along with a novel representation of the action selection process as a tree structure.
    \item A RL framework with multiple actors guided by a single critic network trained with Proximal Policy Optimization (PPO) algorithm for learning to schedule workflows in the cluster.
    \item An end-to-end means of training and deploying the proposed DRL agent in a workflow engine. To the best of our knowledge, this is the first attempt at embedding an intelligent agent in an open source container-native workflow engine.
\end{itemize}

%https://argoproj.github.io/argo-workflows/
%https://argoproj.github.io/argo-workflows/architecture/
%https://argoproj.github.io/argo-workflows/walk-through/

\section{Related Work} 

The use of spot instances for cost optimized workflow scheduling has been studied in a number of studies \cite{zolfaghari2022multi, poola2016enhancing, pham2020evolutionary}. However, the methods proposed in some of these works are associated with bidding strategies \cite{poola2014fault, zolfaghari2022multi} that are of little relevance in current market, since major cloud providers such as Amazon Web Services (AWS) have devised new pricing models that simplifies the purchasing process of spot instances \cite{awsnewpricing}. Accordingly, users are no longer required to analyze historical price trends and employ strategies for determining maximum bid prices. 

In \cite{zhou2019minimizing}, a join cost and makespan optimization algorithm for workflow executions in cloud is proposed. Authors integrated the popular heterogeneous earliest finish time (HEFT) heuristic with fuzzy dominance sort technique for designing the proposed list scheduling algorithm. \cite{belgacem2022multi} also combines the HEFT heuristic with Ant Colony Optimization (ACO) technique for optimizing the same objectives. \cite{zhou2019cost} proposed a makespan and cost aware scheduling technique for hybrid clouds. A combination of Dynamic Voltage and Frequency Scaling (DVFS) and approximate computing is used in \cite{stavrinides2019energy} for energy efficient and cost optimized workflow scheduling in cloud computing environments.   

In \cite{ismayilov2020neural}, authors incorporate artificial neural network with  the NSGA-II algorithm for optimizing a combination of objectives associated with workflow scheduling in cloud computing environments. In \cite{ZhouTPDS2022}, Zhou et. al. proposed optimization framework for HPC applications deployment on clouds in cost-efficient manner. They leveraged cloud spot market resources with the goal of minimizing application cost while ensuring performance constraints.

In \cite{wang2019multi}, a deep Q learning based multi-agent deep reinforcement learning technique is proposed for optimizing cost and makespan of workflow scheduling in cloud. The work models multi-agent collaboration as a Markov game with a correlated equilibrium, so that the makespan and cost agents are not motivated to deviate from the joint distribution in a unilateral manner.
H. Li et. al \cite{li2022weighted} proposed a weighted double deep Q network based reinforcement learning method for cost and makespan optimized workflow scheduling in cloud environments. Scheduling process includes two levels, in the first level a task is selected from amongst all ready tasks. A pointer network is used for efficiently handling the variable length of the input state. In the second level a VM is selected for executing the selected task. A separate sub agent with a separate reward is used for each objective at each level of the scheduling process. 
Y. Qin et. al \cite{qin2020energy} used Q learning for minimizing makespan and energy consumption of workflow executions while adhering to a budget constraint. 

\begin{figure*}[!t]
	\centering
\includegraphics[width=0.75\textwidth, height=5cm]{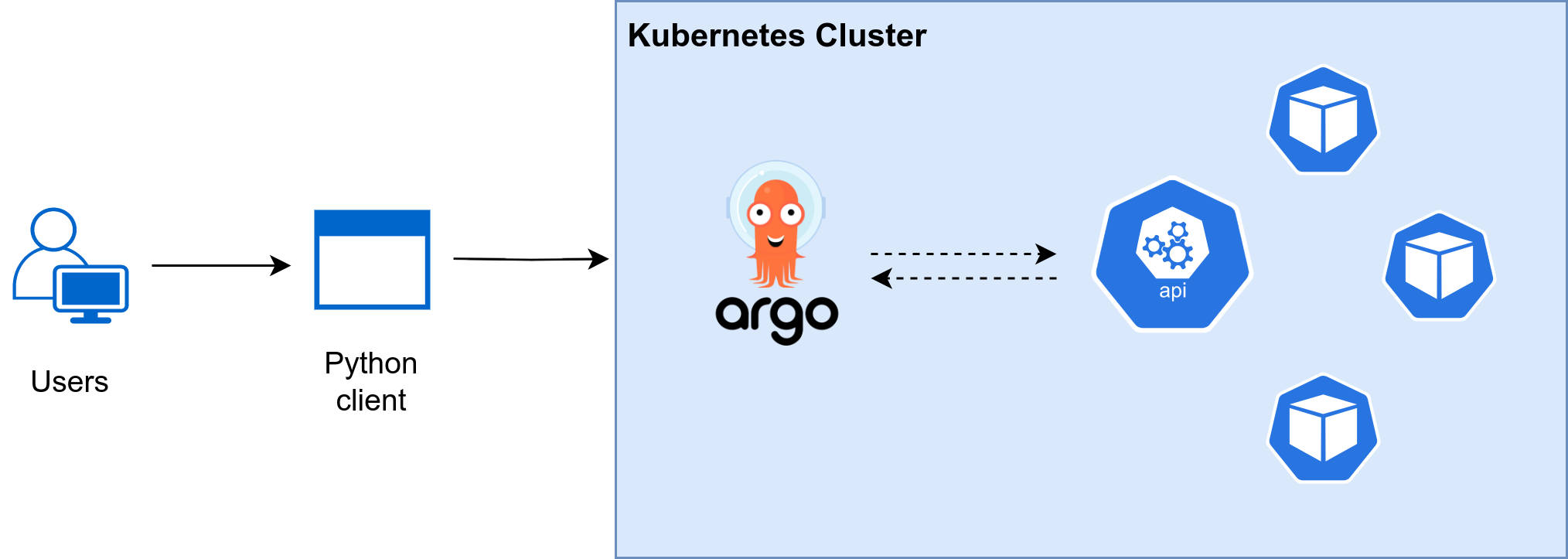}
	\caption{System Architecture}
	\label{fig:high-level-architecture}
\end{figure*}

\section{Problem Formulation}

The objective of the scheduling framework is minimizing the monetary cost of workflow executions, while also minimizing the execution times. The resource requirements in terms of CPU and memory and the dependencies of workflow tasks are included in the submitted workflow specifications. In the workflow specification submitted by users, a workflow is represented by a DAG, $G=(V,E)$ where the nodes, $V=\{v_0, v_1..v_n\}$ of the DAG represent tasks of the workflow, and the edges, $E=\{(v_i,v_j) | v_i,v_j \in V\}$ of the DAG represent precedence constraints between tasks. The computation time of a task, $t_j$ can be represented as:

\begin{equation}
CT(t_j) = \frac{L(t_j)}{F}
\end{equation}

where  $L(t_j)$ is the size of task, $t_j$ and $F$ is the processing rate of the node to which is it assigned.  All the precedence constraints of task, $t_j$ must be satisfied before its execution commences. Accordingly, the execution of all the predecessors must be completed, and the output data required for the execution of $t_j$ must be transmitted to the node in which it is scheduled. If $t_i$ is an immediate predecessor of $t_j$ and the size of data to be transferred from $t_i$ to $t_j$ is $D(t_i,t_j)$, then the total transmission time ($TT$) can be denoted as follows:

\begin{equation}
TT(t_i,t_j) = \frac{D(t_i,t_j)}{B}
\end{equation}

where $B$ is the bandwidth between the execution nodes of $t_i$ and $t_j$. Task execution delay, $TD(t_j)$ primarily depends on the computation time, $CT(t_j)$ of the task, and the maximum data transfer time from predecessor nodes, $\max_{t_i \in pred(t_j)} TT(t_i,t_j)$. The waiting time, $WT(t_j)$ before a task gets scheduled also contributes to total execution delay. Accordingly, $TD(t_j)$ can be represented as: 

\begin{equation}
TD(t_j) = CT(t_j) + WT(t_j) + \max_{t_i \in pred(t_j)} TT(t_i,t_j)
\label{task-execution-time}
\end{equation}

The finish time, $FT(t_j)$ of task, $t_j$ that started execution at time, $ST(t_j)$ can then be expressed as:

\begin{equation}
    FT(t_j) = ST(t_j) + TD(t_j)
\end{equation}

The completion time, $MT$ of a workflow is equivalent to the time at which that last task of the workflow completes execution. It can be denoted as: 

\begin{equation}
MT = \max_{t_j \in T} (FT(t_j)) 
\end{equation}

where T represents the set of all tasks of the workflow.

The computation cost of $t_j$ that executes in a Node with unit cost per second, $UC$ can be represented as:

\begin{equation} \label{eqn:cost-estimate}
    CC(t_j) = CT(t_j) * UC
\end{equation}

The cost of execution, $MC$ of a workflow is equivalent to the sum of execution costs of all tasks, and it can be denoted as follows:

\begin{equation}
    MC = \sum_{t_j \in T} CC(t_j)
\end{equation}

The objective of the scheduling problem is to minimize the cost of workflow executions, and it can be denoted as follows:

\begin{equation}
 \text{Minimize: } \sum_{i=1}^{N} MC_i \\
\end{equation}

where $N$ is the total number of workflows submitted to the system. 

%\textbf{See Tawfiq's paper}

\section{Background and Proposed Approach}

\begin{figure*}[!t]
	\centering
\includegraphics[width=0.7\textwidth, height=10cm]{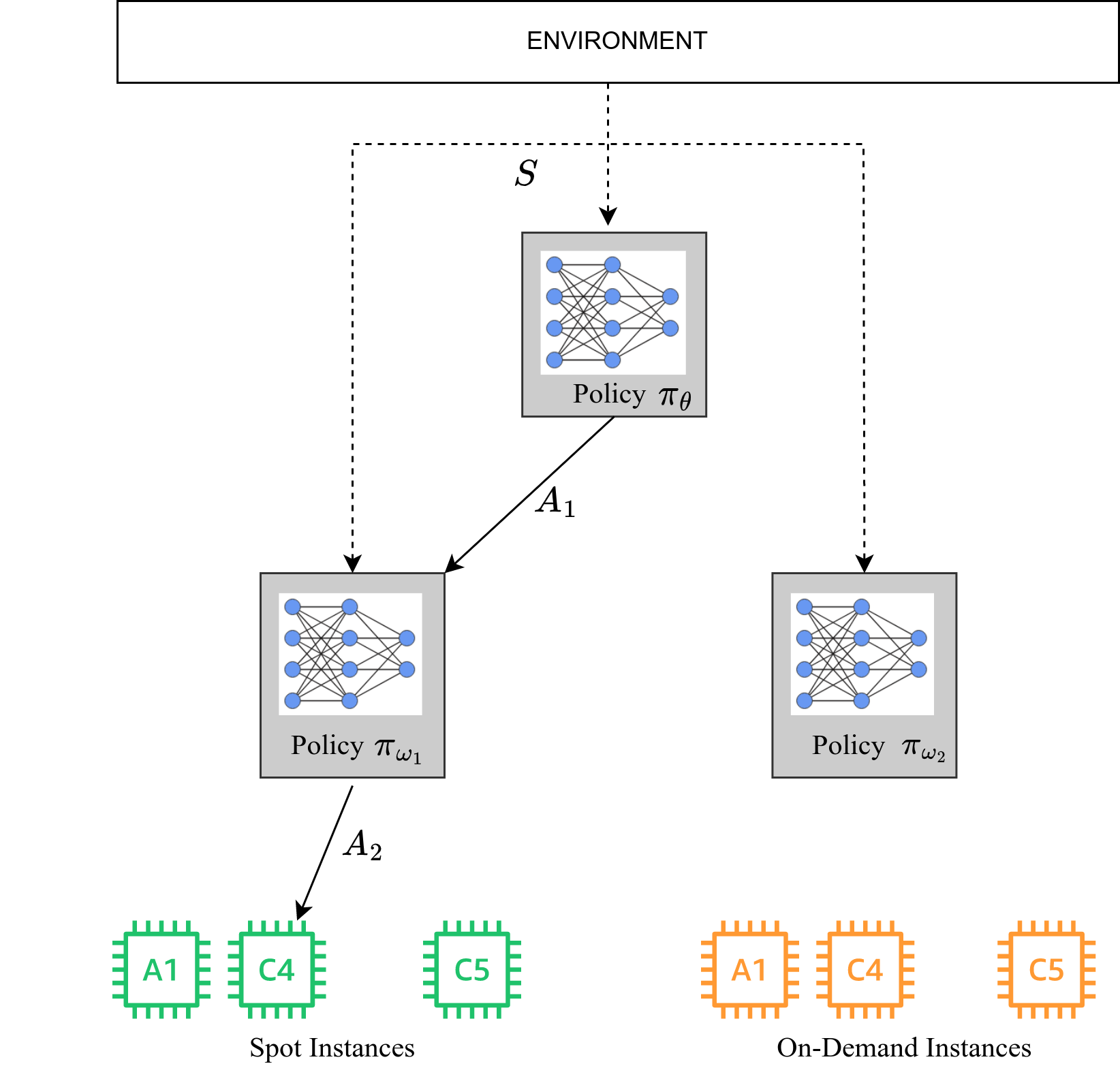}
	\caption{Proposed hierarchical action space and multi-actor DRL model}
	\label{fig:architecture}
\end{figure*}

%The problem formulation presented in the previous section and the RL approach presented in this section can be adapted for a variety of deployment topologies including those that runs across multiple subnets, multiple clouds, availability zones, racks and so on. We have implemented the proposed RL method in a Kubernetes cluster that runs across multiple subnets. Network sub-netting is widely adopted particularly by large organizations for enhancing the efficiency of network routing by eliminating the need for traffic to go through unnecessary routes. This in turn reduces congestion and bottlenecks that results from all network traffic traveling through the entire network.

In this section, we present a background of the popular container orchestration engine Kubernetes and the open-source Argo workflow engine along with details on how the proposed DRL framework is implemented in the Argo Workflow engine that runs atop the Kubernetes cluster. Worker nodes of the Kubernetes cluster are Virtual Machines with different flavors (compute, memory, and storage capacity of VM instances). Argo workflow engine is deployed in the Kubernetes cluster for the management of workflows submitted by users. The scheduler is responsible for selecting the VMs in which the Pods corresponding to each task of the workflow will be scheduled. A high level
architecture of the system is shown in Figure 1. A sequence diagram indicating integration between key components as implemented is shown in Figure 3.

\subsection{Kubernetes}

%Write about Kubernetes basics, about main Kubernetes components, MORE ON SCHEDULER, and how it can be overridden to implement custom policies. 

Kubernetes is a popular open-source container orchestration engine that facilitates containerized applications to be deployed, scaled, and managed in an automated manner. With Kubernetes, containerized workloads can be conveniently deployed and managed in any infrastructure including public clouds and on-site deployments, as well as hybrid combinations of these as required. Workloads can be seamlessly deployed across multi-cloud environments thus enabling the selection of the most appropriate infrastructure for the execution of different parts of the workload. Furthermore, it facilitates the up-scaling and down-scaling of clusters to suit demand variations of applications, which in turn helps reduce costs due to reduced resource wastage. The need for manual intervention is minimized since Kubernetes monitors the health of the deployment and redeploys new containers in the event of a failure to restore operations, and this helps reduce application downtime. Owing to the multitude of benefits offered by Kubernetes, it has become the defacto platform for the deployment and management of containerized workloads. In this work, we extend the capabilities of the default Kubernetes scheduler by incorporating intelligence into it with the use of RL techniques. 

A Kubernetes cluster consists of a set of virtual or physical machines which are referred to as Nodes. The smallest unit deployable in Kubernetes is referred to as a Pod. Pods are hosted by Nodes. A Pod may comprise one or more tightly coupled containers that share storage and network resources, it also contains a specification of how the containers are to be run. The contents of a Pod run in a shared context, and are always located and scheduled together. Pods and Nodes of a Kubernetes cluster are managed by the control plane. It comprises multiple components that work together for managing the cluster. Kube-API server exposes the Kubernetes API that serves as the front end of the Kubernetes control plane. Cluster data are stored in a key-value store termed etcd. The kube-controller-manager runs several controller processes that monitor and regulate the cluster state. Cloud-controller-manager handles cloud-specific control logic. Kube-scheduler is responsible for scheduling unassigned Pods to Nodes for execution. 

%\textbf{ToDo write about k8 scheduler https://kubernetes.io/docs/concepts/scheduling-eviction/kube-scheduler/}

\subsection{Argo Workflow Engine}

Argo workflow engine is an open-source container-native workflow engine that facilitates the orchestration of workflows on Kubernetes. Argo workflows are implemented as a Custom Resource Definition (CRD) in Kubernetes. This enables Argo workflows to be managed using kubectl and they integrate natively with Kubernetes services including secrets, volumes and Role Based Access Control (RBAC).

The workflow engine comprises two main components: the Argo server and the workflow controller. The Argo API is exposed by Argo server and the controller performs workflow reconciliation. In the reconciliation process, the workflows that are queued based on additions and updates to workflows and workflow pods, are processed by a set of worker goroutines. The controller processes one workflow at a time. Both Argo server and controller run in the Argo Namespace. 

Each task of workflow results in the generation of a Pod. Each pod includes three containers. The main container runs the image that the user has configured for the task. The init container is an init container that fetches artifacts and parameters and makes them available to the main container. Wait container performs tasks related to clean up including the saving of artifacts and parameters. 

Argo provides multiple templates for defining workflow specifications and dependencies. For example, a workflow can be defined as a sequence of steps. Alternatively, DAGs can be used for defining a workflow and its dependencies. As this facilitates the representation of complex workflows and parallelism, in this work we have used DAGs for modeling workflows. 

A workflow specification comprises a set of Argo templates, each with an optional input section, an optional output section, and either a list of steps where another template is invoked by each step or a container invocation (leaf template).  The options accepted by the container section of the workflow specification are the same options as the container section of a Pod specification. 

\begin{figure*}[!t]
	\centering
\includegraphics[width=0.7\textwidth,height=11cm]{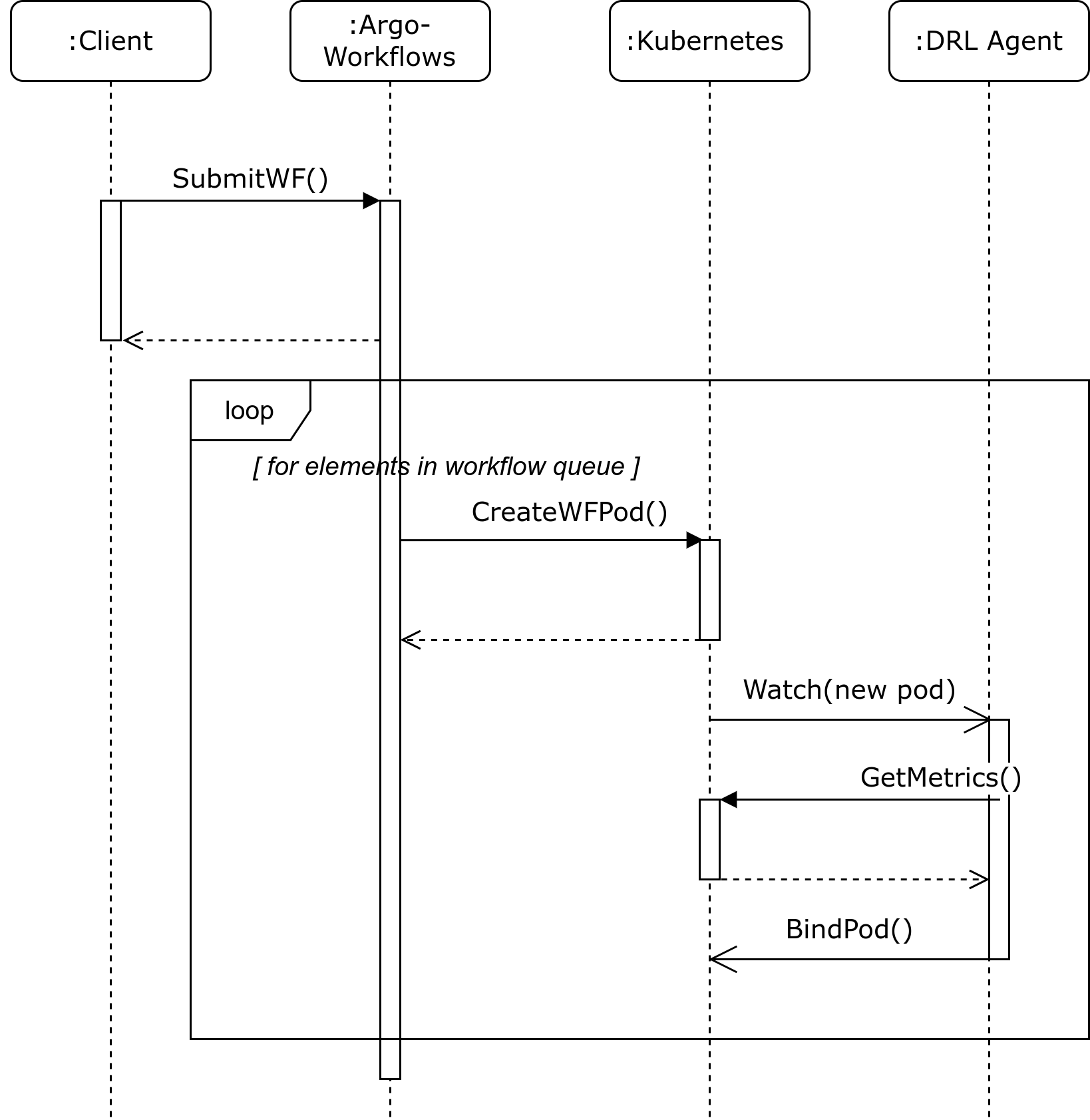}
	\caption{Sequence diagram of DRL based scheduling framework}
	\label{fig:dqn}
\end{figure*}

\subsection{Reinforcement Learning}

In the RL paradigm, an agent learns in a trial-and-error manner by interacting with the environment. The agent receives a \textit{reward}, $r_t$ when it performs an \textit{action}, $a_t$ in a particular \textit{state} $s_t$, and then the \textit{environment} transitions to the next state, $s_{t+1}$. The process repeats until the agent encounters the \textit{terminal state} at which point the \textit{episode} terminates. Markov Decision Process (MDP) can be used for mathematically modeling RL problems. According to the Markov property, it is considered the next state of the environment and the reward received depends solely on the current state and the agent's action in the current state. The cumulative discounted rewards, $G_t$ at any given timestep, $t$ is expressed as:

\begin{equation}
G_t = \sum_{k=0}^{\infty} \gamma^k r_{t+k+1}
\end{equation}

where $\gamma$ is a discount factor and $\gamma \in (0, 1)$. The RL agent operates with the goal of maximizing the expected return, $E[G_t]$ from each state, $s_t$. A \textit{policy}, $\pi(a_t|s_t)$ is a mapping from the current observation of the environment to a probability distribution of the actions that can be taken from the current state. During the training process, a traditional RL agent is required to visit all the states of the problem and store experiences in space-consuming tabular formats. This is a limitation that makes it infeasible to apply the traditional RL paradigm to problems with high dimensional states and action spaces. The integration of Deep Learning with the RL paradigm gave rise to an efficient means of overcoming the aforementioned limitation through the use of neural networks as function approximators for enabling the agent to estimate the value of a state or an action when it encounters a similar circumstance. In the resulting Deep Reinforcement Learning (DRL) paradigm, the policy, $\pi(a_t|s_t)$ is modeled as a parameterized function, $\pi_\theta(a_t|s_t)$ where $\theta$ is an adjustable parameter derived with an RL algorithm. 

In \textit{value based} RL methods, the RL agent attempts to learn a state-value function, $v_{\pi_\theta}(s)$, or a state-action value function, $Q_{\pi_\theta}(s,a)$. As the name implies, the state-value function estimates the value of a state, and it can be expressed in terms of expected return when following a policy $\pi_\theta$ starting from the state, $s$ as shown in Equation \ref{practical-eqn:value-function}. Equation \ref{practical-eqn:action-value-function} indicates the state-action value function which is the expected return when action, $a_t$ is taken at state, $s_t$, and policy, $\pi_\theta$ is followed afterward.

\begin{equation}
v_{\pi_\theta}(s) = E_{\pi_\theta}[G_t | s_t = s] 
\label{practical-eqn:value-function}
\end{equation}

\begin{equation}
Q_{\pi_\theta}(s, a) = E_{\pi_\theta}[G_t | s_t = s, a_t = a] 
\label{practical-eqn:action-value-function}
\end{equation}

In \textit{policy gradient} RL methods, the agent directly learns the policy, $\pi_\theta(a_t|s_t)$. Typically gradient-based techniques on the expectation of returns are used for learning the policy. Equation \ref{practical-eqn:gradient-estimator} indicates the form of the most commonly used gradient estimator. 

\begin{equation}
\hat{g} = \hat{E_t}[\nabla_\theta \ln \pi_\theta (a_t | s_t) \hat{A_t}]
\label{practical-eqn:gradient-estimator}
\end{equation}

where, $\hat{A_t}$ is an estimator of the advantage function at timestep, $t$ and $\pi_\theta$ is a stochastic policy. In an RL algorithm that alternately performs sampling and optimization, the expectation $\hat{E_t}[..]$ indicates the empirical average computed over a batch of samples. For evaluating the performance of the policy, a performance objective the gradient of which is the policy gradient estimator, $\hat{g}$ is  defined. Accordingly, $\hat{g}$ is obtained by differentiating the objective:

\begin{equation}
L^{PG}(\theta) = \hat{E_t}[\ln \pi_\theta (a_t | s_t) \hat{A_t}]
\label{practical-eqn:loss}
\end{equation}

Although multiple rounds of optimizations can be performed on the loss, $L^{PG}(\theta)$ defined in Equation \ref{practical-eqn:loss} using a single trajectory of experience samples, it is not desirable since that could lead to adverse consequences such as policy updates that are destructively large. In order to overcome the aforementioned issue, in Proximal Policy Optimization \cite{SchulmanWDRK17} method, a clipped surrogate objective is used. More specifically, the degree to which new policy, $\pi_\theta(a_t|s_t)$ is allowed to change from old policy, $\pi_{\theta_{old}}(a_t|s_t)$ is restricted by the use of a clip function as indicated in Equation \ref{practical-eqn:ppo-loss}. The clip function, $\text{clip} (r_{t}(\theta), 1 - \epsilon, 1 + \epsilon)\hat{A_t}$ removes the desirability of large policy updates that changes the $r_t(\theta)$ ratio beyond the interval $[1 - \epsilon, 1 + \epsilon]$.

\begin{equation}
\begin{aligned}
    & L^{CLIP}(\theta) = \hat{E_t} [\text{min}(r_{t}(\theta)\hat{A_t}, \text{clip} (r_{t}(\theta), 1 - \epsilon, 1 + \epsilon)\hat{A_t}] \\
    & \text{where } r_{t}(\theta) = \frac{ \pi_\theta(a_t|s_t) } {\pi_{\theta_\text{old}}(a_t|s_t)}
    \end{aligned}
    \label{practical-eqn:ppo-loss}
\end{equation}

Actor-critic is a branch of RL algorithms that combines the advantages of value-based methods and policy gradient RL methods. The actor is the policy that outputs a probability distribution over the actions that can be taken in the current state, and the critic is the value function approximator that evaluates the actions taken by the actor as per the policy. 

%\textbf{ToDo write about actor and critic updates here}

\begin{algorithm}[!t]
	\caption{Actor-Critic based Scheduling Framework with PPO}\label{algo:dqn}
	\begin{algorithmic}[1]
		\State Initialize actor networks and critic network with random weights
		\State Initialize the training parameters:  $\alpha, \beta, \gamma$
		\For{episode = 1 to $N$}
		\State Reset the environment
		\For{step = 1 to $T$}
		\indentedState{%
		Input the state of the environment to actor networks}
		\State Select action $a_1$ from ${\pi_\theta}$
		\indentedState{%
		Select action $a_2$ from $\pi_{\omega_i}$} 
		\indentedState{%
		Execute the combined action $(a_1, a_2)$ and observe the corresponding reward $r_t$ and next state of the system $s_{t+1}$}
		\indentedState{%
		Store the most recent transition $(s_t,a_t,r_t,s_{t+1})$ in memory $D$}
		\EndFor
		\State Compute advantage estimates $\hat{A_1}$ to $\hat{A_T}$
		\For{j = 1 to $K$}
		\indentedState{%
		Randomly sample a mini-batch of samples of size $S$ from $D$}
		\For{p = 1 to $S$}
		\indentedState{%
		Update critic network: \\ $\sigma \gets \sigma + \beta \delta_t \nabla v_\pi(s_t|\sigma)$}
		\indentedState{%
		Update first actor network: \\ $ \theta \gets \theta + \alpha \hat{A_p} \nabla \ln \pi(a_1| s, \theta) $}
		\indentedState{%
		Update second actor network: \\ $ \omega \gets \omega + \gamma \hat{A_p} \nabla \ln \pi(a_2| s, \omega) $}
		\EndFor
		\EndFor
		\State Clear memory $D$
		\EndFor
		\Return
	\end{algorithmic}
\end{algorithm}

\subsection{Proposed RL Framework}

 As previously discussed, the default kube-scheduler takes multiple factors into account in formulating scheduling decisions including resource requirements and constraints, specifications of affinity and anti-affinity, deadlines, and interference caused by co-located workloads. These policies need to be pre-defined and may suffer from the general limitations of heuristic scheduling techniques. In this work, we override the default behavior and incorporate intelligence into the scheduler by training a DRL agent to select appropriate scheduling decisions with the objective of achieving a desired goal.

 \subsubsection{Agent Environment}

 The problem of scheduling workflows in a cloud cluster can be simplified by formulating it as a dependent task-scheduling problem. In the Argo workflow engine, pods corresponding to independent tasks are scheduled directly in the cluster for execution, while the tasks with dependencies are not scheduled until the parent tasks have completed execution. Whenever the workflow scheduler (RL agent), discovers a pod that is not assigned to a node, it takes the current state of the environment as input and outputs the most desirable node for task execution based on the trained policy. The environment then transitions to the next state. Accordingly, the timesteps of the proposed RL model are discrete and event-driven. The state, action, and reward of the RL model are designed as follows:

 \textit{State Space:} State of the environment comprises of total CPU and Memory requirements of the task, and nodes together with the estimated waiting time at each node based on the number of pods executing in each node.

\textit{Action Space:} Compared to the problem of scheduling tasks in a cluster comprising nodes from the same cloud data center, scheduling tasks in a multi-cloud cluster is more challenging since resource capacities and cost are not the only factors that differentiate nodes. In such scenarios, the intercloud communication delay is an important factor that needs to be factored into the formulation of scheduling decisions. This requirement is further heightened in workflow scheduling due to the presence of data dependencies among tasks that may result in costly data transfers if communication costs among nodes from different clouds are ignored. 

% As opposed to the problem of independent task scheduling, dependent task scheduling is more complex since the data dependencies need to be taken into account in the formulation of scheduling decisions. 

% The most straightforward design of the action space is to include all nodes from all clouds as actions in a flat action space. 

 In the most straightforward design of the action space, the action of selecting any one of the nodes in the multi-cloud cluster can be represented together in a flat action space. In this approach, the burden of distinguishing nodes from different clouds lies with the DRL agent.  Although the agent may eventually manage to learn the presence of nodes from multiple clouds based on rewards and thereby develop an internal representation of the multi-cloud composition of the nodes, it will inevitably reduce the training efficiency of the agent. Furthermore, as the size of the cluster grows, flat action spaces are more prone to the problem of the 'curse of dimensionality'. 

 In order to efficiently overcome the aforementioned challenges, we have designed the action space considering a logical organization of cluster. In the logical organization, nodes from different pricing categories are grouped together as shown in Figure \ref{fig:architecture}. Accordingly, we define a hierarchical action space for the problem as follows:

     \begin{equation}
         A = \{(a_1, a_2) | a_1 \in \{\pi_{\omega_1}, \pi_{\omega_2}\} \And a_2 \in \{1,2,...,N_{a_1}\}\}
    \end{equation}
    
    where $N_{a_1}$ is the total number of nodes in the cluster that belong to the group given by action $a_1$. The action, $a_1$ corresponds to the selection of a node group, and the action, $a_2$ corresponds to the selection of a node from the group. An action at each timestep then corresponds to the joint action $(a_1,a_2)$.

 \textit{Reward:} Reward is the estimated cost of execution at the allocated node computed with Equation \ref{eqn:cost-estimate}.

\subsubsection{Multi-Actor RL Algorithm}

The hierarchical action space described above can be represented as the tree structure in Figure \ref{fig:architecture}. Each level of the tree corresponds to an action selection sub-problem. The first level of the tree represents the sub-problem of selecting a node group and the second level represents the  sub-problem of selecting a node. We then adopt the hybrid actor-critic technique presented in \cite{fan2019hybrid} for selecting joint actions from the hierarchical action space. Different from a traditional actor-critic algorithm which contains a single actor-network and a single critic network, in the proposed architecture multiple parallel actor networks are guided by a common critic network. 

As shown in Figure \ref{fig:architecture} each action-selection sub-problem is handled by a separate actor network. Accordingly, one actor network learns a stochastic policy for selecting a node group. For each of the node groups, a separate actor network learns a stochastic policy for selecting a node from the respective node group. The critic network estimates the state value function, $V(s)$. The advantage function provided by the critic network is used for updating the stochastic policies. Actor networks are separately updated at each timestep by their respective update rules. We used the PPO method for updating the networks. Algorithm 1 summarizes the steps included in the
training process of the DRL agent. 

%The action, $a_1$ corresponds to the first level sub-problem of selecting the node group, and the action, $a_2$ corresponds to the second level sub-problem of selecting a node from the group.

% And the use of multiple actors, one actor at the top of the hierarchy determines the cloud, and then one for each cloud guided through one critic \\

% Physically - a single cluster, but it contains nodes from multiple clouds
% Logically - can be separate into sets of nodes from multiple clouds

% This leads to a hierarchical organization of the multi-cloud scheduling problem, which in turn can be represented as a tree structure. The first level of the tree corresponds to the selection of a cloud and the second level represents the selection of a node.

\begin{table*}
  \centering
  \renewcommand{\arraystretch}{1.2}
  \begin{tabular}{|l|c|c|c|c|c|c|c|}
    \hline
    \textbf{Instance Type} & \textbf{CPU Cores} &  \textbf{Memory(GB)} &  \multicolumn{2}{c|}{\textbf{Quantity}} &
    \multicolumn{2}{c|}{\textbf{Price}}\\
    \cline{4-7}
    & & &  \textbf{Spot} & \textbf{On-demand} & \textbf{Spot} & \textbf{On-demand} \\

    \hline
     t4g.large & 2 & 8 & 2 & 2 & \$0.033/h & \$0.0672/h  \\
     t4g.xlarge & 4 & 16 & 3 & 2 & \$0.0857/h & \$0.1344/h  \\
     t4g.2xlarge & 8 & 32 & 1 & 1 & \$0.1589/h & \$0.2688/h  \\
     \hline
  \end{tabular}
  \\
  \caption{Resource configurations of Kubernetes cluster}
  \label{table:kubernetes-specs}
\end{table*}

\begin{figure*}[!t]
    \centering
\subfloat[Execution cost]{\includegraphics[width=0.65\columnwidth]{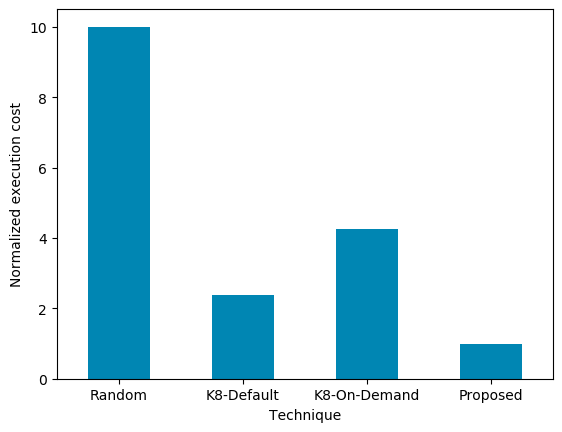}}
\hfill
\subfloat[Execution time]{\includegraphics[width=0.65\columnwidth]{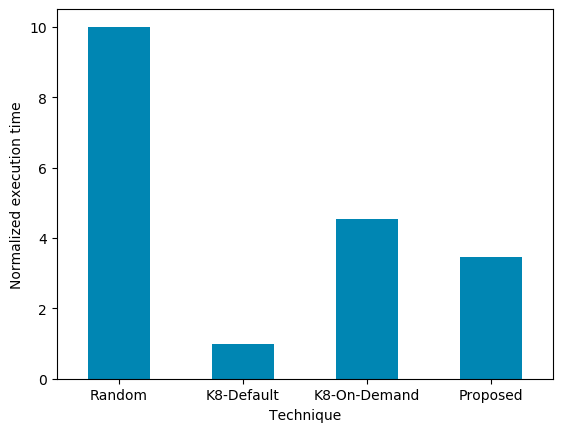}}
\hfill
\subfloat[Execution interruptions]{\includegraphics[width=0.65\columnwidth]{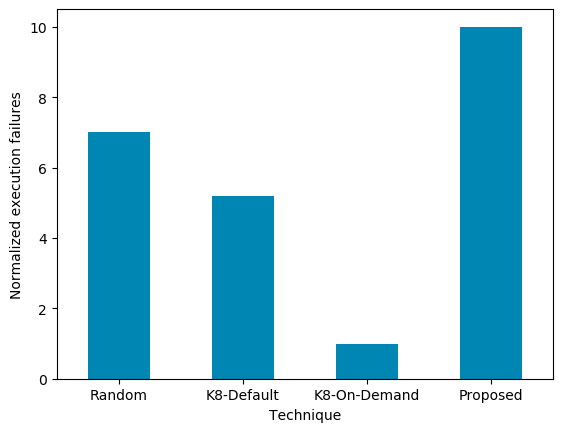}}
	\caption{Comparison of performance of scheduling algorithms on an experimental dataset}
\label{fig:cost-perf}
   \end{figure*}
   
\section{Performance Evaluation}

In this section we present the details of the experimental testbed used for evaluating the proposed DRL framework along with the results of the performance evaluation. 

\subsection{Experimental testbed}

The resource configurations and composition of the Kubernetes cluster is shown in Table \ref{table:kubernetes-specs}. Argo workflow engine is installed in the cluster in a separate namespace. A Python client that communicates with Argo API server was developed for submitting workflows and querying about the execution statistics of workflows. 

\subsection{Experimental dataset}

The experimental dataset comprises of a set of Map-Reduce workflows. Each map task performs a CPU intensive parallelizable computation that involves finding the sum of the square-roots of numbers in a given input range. Experiments were conducted at different arrival rates drawn from a uniform distribution. Experiments were also conducted at different task sizes and parallelism levels.
% Experiments were conducted at different arrival rates sampled from a Poisson distribution. 

\subsection{DRL Scheduler Implementation}

The Argo workflow engine uses the default Kubernetes scheduler for allocating tasks (i.e. pods) to nodes. We have overridden it with a DRL agent trained according to the proposed DRL framework. The configurations of all test workflows were updated such that they are scheduled with the custom DRL scheduler instead of the default scheduler. Keras library \cite{chollet2018keras} was used for developing the DRL framework. 

Kubernetes metrics server collects resource metrics of the underlying nodes from Kubelets and shares it with the Kubernetes API server via the Metrics API. Therefore, by querying the Kubernetes API server we were able to retrieve
near real-time CPU and Memory usages of the nodes, which were required to formulate the state space composition that needs to be provided as the state of the environment to the agent at each timestep of the episode. At the end of each episode, the python client queries the Argo API server for retrieving the execution statistics of workflows including resource times, start and end times of workflows and success rates which are then used for computing the resource usages and associated costs. The client also queries Kubernetes API server for retrieving node metrics that is required for computing the up times of nodes. 

\subsection{Comparison Algorithms}

The performance of the proposed DRL algorithm was compared against three scheduling policies. \textbf{Random} policy allocates tasks to nodes in a Random manner, and is completely agnostic to pricing as well as other resource utilization levels of the cluster. \textbf{K8-Default} refers to the default scheduling policy of Kubernetes cluster.  \textbf{On-Demand} is a policy that uses Kubernetes default scheduler but the selection is limited to the on-demand instances. 

\subsection{Experimental Results}

Figure \ref{fig:cost-perf}a shows the performance of the algorithms on the experimental dataset with respect to monetary cost of workflow executions. Random algorithm has incurred the highest cost owing to the fact that it distributes tasks across multiple instances without trying to optimize resource utilization or cost. In comparison the Kubernetes scheduler exhibits much better cost savings. By default, it is designed to select the most appropriate node through a node filtering and scoring process. In the filtering phase, nodes that are feasible for executing the pod are selected, and then they are ranked according to a scoring process. Based on the outcome of the filtering and scoring process the most appropriate node for pod execution is selected. Clearly, this process has resulted in much better resource efficiency and thereby cost savings in comparison to random allocation. As expected, K8-On-Demand method has incurred a higher cost than the default policy since it is only allowed to make a selection from amongst the on-demand instances which have a higher unit cost. The proposed method has resulted in the highest cost savings. The significant reduction in cost is due to the intelligent cost aware allocation of pods among the instances in the cluster.

Figure \ref{fig:cost-perf}b shows a comparison of the execution times of workflows scheduled with different algorithms. Again, the highest amount of time is taken by Random algorithm. K8-On-Demand has resulted in higher execution times compared to k8-Default due to the limited selection of instances available for scheduling. K8-Default has resulted in the least execution time since it distributes pods amongst multiple high scoring nodes, without considering the respective unit cost differences. Proposed algorithm has incurred slightly higher cost since it's favoring nodes that are of low cost which leads to more pods being assigned to the same nodes, hence resulting in increased execution times. This is expected since instances with more vCPUS are more expensive, which results in a trade-off between execution time and cost.

Figure \ref{fig:cost-perf}c shows the number of execution failures. Execution failures in the experimental context are solely due to the interruption of spot instances which leads to workflows timing out and thereby failing to complete. As expected the proposed algorithm results is the highest failures since it is 
favoring spot instances for task executions, and the spot instances are subjected to interruptions. This is a known trade-off associated with the use of spot instances, therefore it is important to restrict the use of spot instances for failure tolerant workflows. 

\section{Conclusions}

In this work, we designed a DRL technique for cost-optimized workflow scheduling in Cloud environments by the intelligent use of spot and on-demand instances. We then designed and implemented an end-to-end system for integrating and training the DRL agent in the container-native Argo workflow engine that runs atop Kubernetes. As evidenced by the results of the experiments, higher cost savings can be achieved by overriding the default schedulers with intelligent cost-optimized scheduling policies.

\bibliographystyle{ieeetr}
\bibliography{Bibliography}

\begin{IEEEbiography}[{\includegraphics[width=1in,height=1.25in,clip,keepaspectratio]{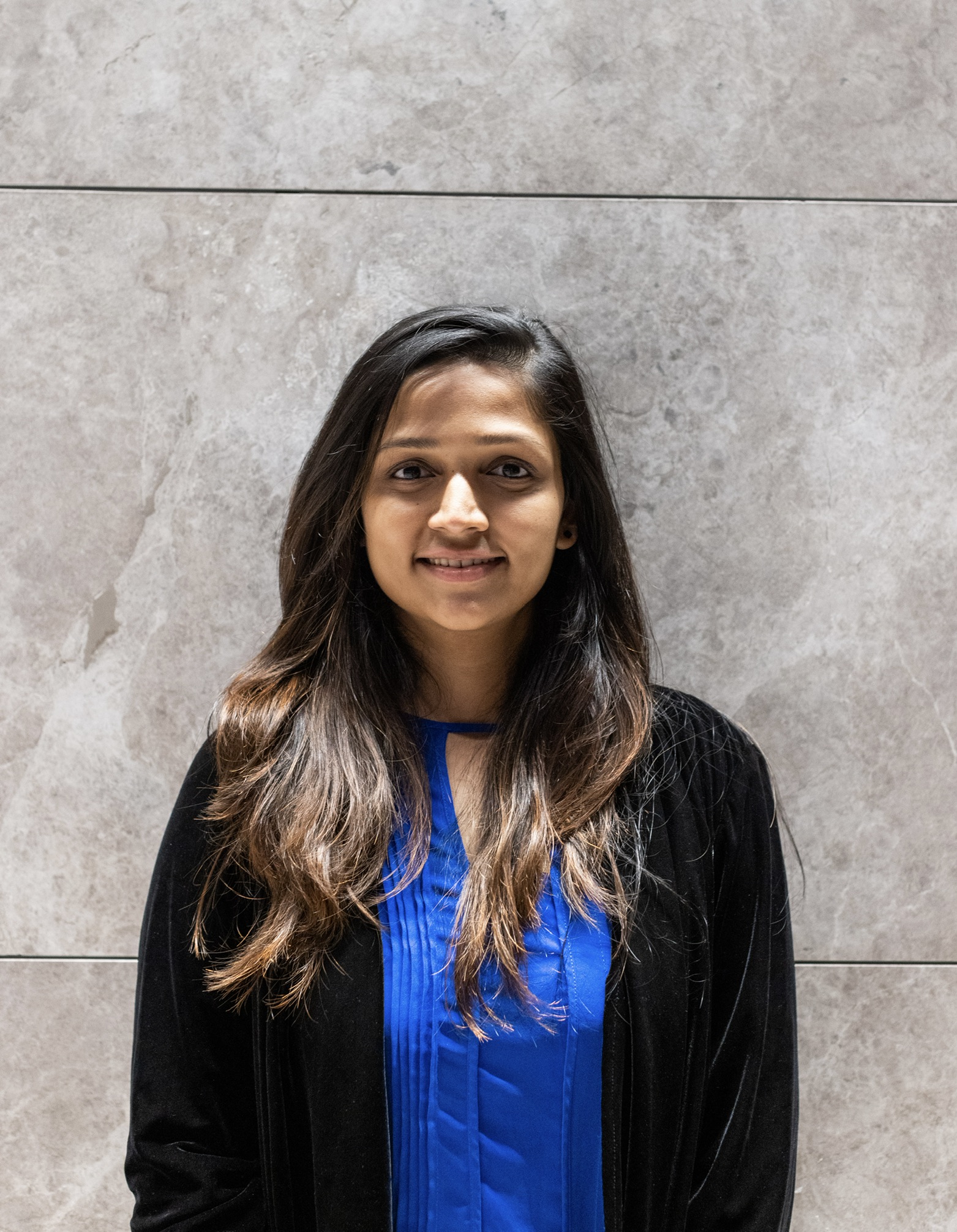}}]{Amanda Jayanetti} is currently working toward the PhD degree at the Cloud Computing and Distributed Systems (CLOUDS) Laboratory, Department of Computing and Information Systems, the University of Melbourne, Australia. Her research interests include Artificial Intelligence (AI), Cloud Computing and Edge Computing. Her current research focuses on harnessing the capabilities of Artificial Intelligence (AI) techniques for enhancing the performance of cloud and edge computing environments. \end{IEEEbiography}

\begin{IEEEbiography}[{\includegraphics[width=1in,height=1.25in,clip,keepaspectratio]{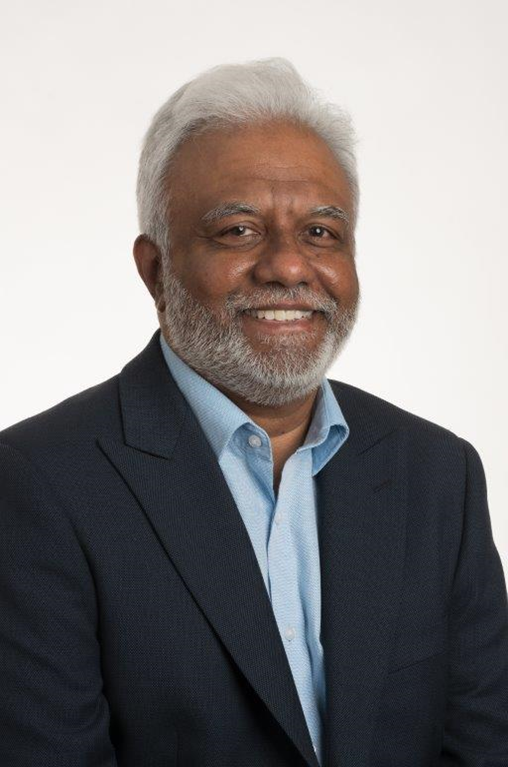}}]{Saman Halgamuge} Fellow of IEEE, received the B.Sc. Engineering degree in Electronics and Telecommunication from the University of Moratuwa, Sri Lanka, and the Dipl.-Ing and Ph.D. degrees in data engineering from the Technical University of Darmstadt, Germany. He is currently a Professor of the Department of Mechanical Engineering of the School of Electrical Mechanical and Infrastructure Engineering, The
University of Melbourne. He is listed as a top 2\% most cited researcher for AI and Image Processing in the Stanford database. He was a distinguished Lecturer of IEEE Computational Intelligence Society (2018-21). He supervised 50 PhD students and 16 postdocs in Australia to completion. His research is funded by Australian Research Council, National Health and Medical Research Council, US DoD Biomedical Research program and International industry. His previous leadership roles include Head, School of Engineering at Australian National University and Associate Dean of the Engineering and IT Faculty of University of Melbourne.\end{IEEEbiography}

\begin{IEEEbiography}[{\includegraphics[width=1in,height=1.25in,clip,keepaspectratio]{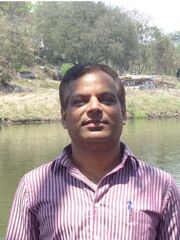}}]{Rajkumar Buyya} is a Redmond Barry Distinguished Professor and Director of the Cloud Computing and Distributed Systems (CLOUDS) Laboratory at the University of Melbourne, Australia. He has authored over 800 publications and seven text books including “Mastering Cloud Computing” published by McGraw Hill, China Machine Press, and Morgan Kaufmann for Indian, Chinese and international markets respectively. He is one of the highly cited authors in computer science and software engineering worldwide (h-index=168, g-index=369, 150,900+ citations). \end{IEEEbiography}

\vfill

\end{document}